\documentclass[reprint, twocolumn, amsmath, amssymb, showpacs, superscriptaddress, aps, prl]{revtex4-2}

\usepackage{microtype}
\usepackage{graphicx}
\usepackage{dcolumn}
\usepackage{bm}
\usepackage[colorlinks=true,
						linkcolor=blue,
						urlcolor=blue,
						citecolor=blue,
						bookmarks=true,
						pdfborder={0 0 0}]{hyperref}
\usepackage{multirow}
\usepackage[normalem]{ulem}
\usepackage{xcolor}
\usepackage[export]{adjustbox}

\usepackage{url}

\begin{document}
\title{Evolutionary game model of group choice dilemmas on hypergraphs
}
\author{Andrea Civilini}
\affiliation{School of Mathematical Sciences, Queen Mary University of London, London E1 4NS, United Kingdom}
\author{Nejat Anbarci}
\affiliation{Department of Economics and Finance, Durham University, Durham DH1 3LB, United Kingdom}
\author{Vito Latora}
\affiliation{School of Mathematical Sciences, Queen Mary University of London, London E1 4NS, United Kingdom}
\affiliation{Dipartimento di Fisica ed Astronomia, Universit\`a di Catania and INFN, Catania I-95123, Italy}
\affiliation{Complexity Science Hub Vienna, A-1080 Vienna, Austria}

\begin{abstract}
We introduce an evolutionary game on hypergraphs in which decisions between a risky alternative and a safe one are taken in social groups of different sizes. The model naturally reproduces choice shifts, namely the differences between the preference of individual  decision makers and the consensual choice of a group,  that have been empirically observed in choice dilemmas. In particular, a deviation from the Nash equilibrium towards the risky strategy occurs when the dynamics takes place on heterogeneous hypergraphs. These results can explain the emergence of irrational herding and radical behaviours in social groups.

\end{abstract}

\maketitle

{\it Choice dilemmas} describe the most general situations in which decision makers are faced with an alternative between two possibilities: a riskier strategy, that either brings a high reward, with a probability $w_p$, or a low one, with probability $1-w_p$, and a safer strategy with an intermediate reward \cite{stoner1961comparison, kogan1967risky, eliaz2006choice}.
Expected Utility Theory and its strategic version Game Theory
assume that rational decision makers always act to maximize their private utilities/payoffs \cite{von1944theory}. However, over the years a series of empirical evidences in contrast with the theoretical predictions have clearly pointed out the descriptive limitations of these theories and have 
undermined their very fundamental assumptions. 
Alternative explanations of these discrepancies are based on ad hoc behavioural mechanisms in the decision making process \cite{machina1987choice, starmer_nonEUT_2000, DAVIS1992, stoner1961comparison, Edward1955, davis1974social, eliaz2006choice, garling2009psychology, bikhchandani2000herd, sias2004institutional, lux1995herd} such as loss aversion \cite{starmer_nonEUT_2000},  risk diffusion \cite{wallach_group_1962, wallach_diffusion_1964},  rational conformity \cite{kameda_herding_2015}, social facilitation \cite{zajonc_social_1970} and 
{group polarization \cite{kogan1967risky, laughlin_social_1982, DAVIS1992}, i.e. individual opinion polarization through social interaction processes \cite{mas2013_differentiation, shin2010tipping, baumann_echochambers_2020, wang_echo_chambers_2020} such as persuasive argumentation and social comparison \cite{BURNSTEIN1977315}.}
In particular, two anomalies of great practical relevance arise when decision makers interact in social groups, influencing each other. The first anomaly is the herd behaviour leading to irrational outcomes, observed for example during financial bubbles, where decision makers can follow the opinion of others, apparently disregarding their own interests \cite{garling2009psychology, bikhchandani2000herd, sias2004institutional, lux1995herd}. The second anomaly has been empirically observed when decision makers are organized in groups and decide collectively their strategy \cite{kogan1967risky, DAVIS1992, eliaz2006choice, Davis1973GroupDA} (a precious source of insights and experimental data is offered by the studies on trial juries conducted in the Seventies \cite{davis1975decision, DAVIS1977, Nemeth1977, bray1978authoritarianism, penrod1980computer}, see Supplemental Material, SM~\footnotemark[1]). 
In this context a {\it choice shift} effect emerges, in which the average opinion of individual decision makers is exacerbated when they act in group \cite{stoner1961comparison, DAVIS1992, laughlin_social_1982}. 
All the existing theories for herd behaviour and choice shift have some major drawbacks. 
First of all, no existing theory is able to reproduce both anomalies at the same time.
Secondly, existing models consider isolated social groups of fixed size, neglecting the presence of groups of heterogeneous sizes and the nested hierarchical structure of real social systems, which are known to play a key role in the emergence of critical phenomena \cite{barrat2008dynamical, pastor2015epidemic, dorogovtsev2008critical}.
Finally, the existing models focus on either one or the other of the two main aspects of group decision: information spreading  \cite{zajonc_social_1970, kameda_herding_2015, Lahno_2015_peereffect} and the aggregation of preferences 
\cite{eliaz2006choice, Laughlin1986socialcombination, DAVIS1992}.

In this Letter, we propose to use evolutionary game theory on hypergraphs to model group choice dilemmas.
Higher-order interactions have recently been shown to influence dynamical processes 
\cite{BATTISTON2020}, including social 
contagion \cite{iacopini_simplicial_2019}
and evolutionary games 
\cite{Traulsen_multilevel2006, alvarez-rodriguez_evolutionary_2021}. Our model can be seen as a generalization to higher-order interactions of anti-coordination pairwise games, such as the game of Chicken \cite{gameofchicken_1966}, in which 
the decision makers are organized in groups of different sizes and each decision maker can participate to a variable number of groups.
Differently from all previous models, our game dynamics describes, at same time, how opinions spread according to an imitation process and how they are aggregated by the members of a group to determine the group strategy.
We show that, due to this, group 
choice shifts emerge naturally in our model, and we explain how they depend on the mechanisms of preference aggregation and on the structure of the hypergraph. In particular, when implemented on heterogeneously structured populations, our model predicts the spontaneous emergence of irrational herd behaviours towards the riskier strategy.

{\bf Model.~~} 
A population of $M$ interacting decision makers is modelled as the nodes of a hypergraph, whose $N$ hyperedges describe the interactions in groups of two or more agents  \cite{BATTISTON2020, ESTRADAhypergraph}. 
The hypergraph can be represented by a $M \times N$ adjacency matrix $A$, whose entry $a^{(g)}_i$ is equal to $1$ if the agent $i$ is in group $g$, or is zero otherwise. 
The hyperdegree of agent $i$, $ k_i = \sum_{g \in N} a_i^{(g)} $, and the size of group $g$,
$q^{(g)} = \sum_{i \in M} a_i^{(g)}$, represent respectively the number of groups in which agent $i$ takes part, and the number of members of group $g$.
Let $Q(q)$ and $K(k)$ be the probability distributions of group sizes and agent hyperdegrees respectively \cite{bipartitegraphs}. 
In particular, we focus on power-law distributions $Q(q)\sim q^{-\lambda}$ and
$K(k) \sim k^{-\nu}$.  In fact, many real-world group sizes are power-law distributed: scientific teams \cite{newmancoauthornet, milojevic_teamformation}, firms \cite{Axtell_firm_sizes}, human settlements/cities \cite{gabaix2004evolution} and also groups of animals \cite{powerlaw_antelope, niwa2003power}.
Moreover, it has been shown that also the number of social activities/groups in which a person is involved (i.e. the hyperdegree)  follows long-tailed distributions \cite{newmancoauthornet, muchnikpowerlawhumanactivities, perra_activity_2012}.
Given the probability of success of a risky activity $0 \leq w_p \leq 1$, each agent can be in one of two states: $\textbf{s}$ (safe) or $\textbf{t}$ (tempted). If in state $\textbf{s}$, the agent prefers not to participate to the risky activity. Instead if in state $\textbf{t}$, agent is willing to share with the other group members the cost $C$ to attempt this risky strategy. 
The strategy of a group, namely the decision to participate to the risky activity (group strategy $\textbf{T}$) or not (strategy $\textbf{S}$),  depends on the states of the group members and on the way in which these states/personal opinions are aggregated. We have considered different \emph{group decision schemes}, 
i.e. rules of opinion aggregation, ranging from simple majority (where the group adopts the strategy preferred by at least half of the group members + 1), to two-third majority, and  proportionality \cite{hastie2005_beautyofmajority}. 
If we indicate with $C$ the cost for the group to adopt a risky strategy, and with $T$ and $S$ the rewards, respectively in the case of success or failure
of the risky activity, then we assume the following group payoffs $\pi^{(g)}$ :
\begin{align}
\label{eq:payoff}
\pi^{(g)}(\textbf{S}) &= S \\
\pi^{(g)}(\textbf{T}) &= \begin{cases}
            W := T-C \text{, with probability $w_p $} \nonumber \\
            L := S-C \text{, with probability $1-w_p$}
            \end{cases}
\end{align}
associated with the group strategies. 
Given $ 0 < C \leq S < T$, it follows that $L < S < W$ and then we are in a classical choice dilemma scenario, where the risky choice $\textbf{T}$ brings a high payoff $W$ (with probability $w_p$) or a low payoff $L$ (with $1-w_p$),
while the safe action $\textbf{S}$ guarantees an intermediate payoff $S$.
Let us define as $z=N_T/N$ the fraction of groups with strategy $\textbf{T}$.
In our model at each time step the $zN$ groups with strategy $\textbf{T}$ are in competition for a fixed share $W$ of the total reward $\theta N W$,
with $0<\theta < 1$. As a consequence, the probability of success $w_p$ of a risky activity is modelled as a non-decreasing function of the total number of shares $\theta N$ per group with strategy $\textbf{T}$: $w_p=f\left(\frac{\theta }{z}\right)$. Moreover, in our model the successful groups are selected uniformly at random among the $zN$ groups with strategy $\textbf{T}$.
Hence, the higher is $z$, the fraction of player with strategy \textbf{T}, the lower is the chance of high reward (i.e. actions \textbf{T} are strategic substitutes \cite{bulow_1985}). We show in the SM~\footnotemark[1] that, when played by 
two groups of equal size, our game is equivalent to the pairwise game of Chicken 
\cite{Bruns2015NamesFG}.
To keep the model as general as possible, we allow the cost $C$ to be a non-increasing function $C(r^{(g)}) = (r^{(g)})^{-1}$ of a group \emph{resource} $r^{(g)} \in \mathbb{R}^{+}$. 
The idea is that the more resource a group has, the lower the cost for attempting the risky activity and the risk perception are \cite{wallach_group_1962, wallach_diffusion_1964}.
To define the resource of a group, we assume that each agent $i$ is given an individual resource $r_i = r_{\rm min} k_i$ that the agent splits equally among the $k_i$ groups to which it participates. 
This is a realistic assumption, as the greater is the resource of an agent the larger will be on average the number of activities it is involved \cite{campbell_2019_rich_portfolio}. Moreover, this means that agent $i$ invests in each of its $k_i$ activities an amount of resource $r_{i}^{(g)} = r_{\rm min} $. 
Consequently, the total resource $r^{(g)}$ of a group $g$ is a function of its size $q^{(g)}$:
\begin{equation} \label{eq:group_resource_beta}
r^{(g)}=  \left( \sum_{i\in g } r_i^{(g)} \right)^{\beta} = \left( r_{\rm min} q^{(g)}\right)^{\beta}   
\end{equation}
where the exponent $\beta \geq 0$ takes into account possible nonlinear synergistic effects raising from the interaction among group members \cite{Bettencourt7301}. 
In particular, for $\beta > 1$ the interaction among agents leads to a superlinear scaling of the group resource with the group size.
To determine the payoff of the individual agents, we simply assume that the payoff of a group in Eq.~\eqref{eq:payoff} is equally shared among its group members.
Then, the payoff of agent $i$ is defined as the sum of the returns from all the groups in which it is involved: 
$    \pi_i =\sum_{g|i\in g} 
    {\pi^{(g)}} /        {\left(q^{(g)}\right)^{\zeta}}
    $.
The exponent $0 \leq \zeta \leq 1$ allows to tune the way in which group members benefit from the group payoff. For example, in contexts where the payoff represents a material or countable quantity (e.g. a cash prize), $\zeta = 1$ implies that the group payoff is equally split among the group members and the total group payoff is conserved. In the limit $\zeta = 0$ instead each group member earns the full payoff coming from the group~\footnotemark[2].
In the \emph{evolutionary dynamics} of our model, we assume perfect rationality of the agents, meaning that the agent states are updated in time according to a stochastic dynamics where each agent tries to imitate the fittest neighbour \cite{gamedynamicstextbook}.
Namely, at each time step, a \emph{focal} individual $i$ is selected at random in the population. A second individual $j$, that we call \emph{reference}, is randomly selected among the co-members of the focal individual.
Then, the probability $p_{i j}$ for the decision maker $i$ to adopt the state of the decision maker $j$  is defined as a growing function of the payoffs difference between the two individuals: $p_{i j} =  g ( \pi_j - \pi_i)$.
The \emph{co-membership network} we consider is obtained as the projection of the hypergraph on the set of nodes representing group members (two nodes are linked if they are co-members in at least one group). 
We denote as $h_i$ the degree of node $i$ on the co-membership network, i.e. the number of co-members of agent $i$, and as $H(h)$ its distribution (see SM~\footnotemark[1]).
In particular, we verified that for realistic heterogeneously distributed group sizes and hyperdegrees, the resulting co-membership distribution shows power-law tail $H(h)\sim h^{-\gamma}$, as the one observed in many relevant real-world systems \cite{newmancoauthornet, milojevic_teamformation, perra_activity_2012}.

{\bf Results.}
We have investigated how the presence of groups affects decision making through a series of numerical simulations of the model dynamics performed using the quasi-stationary (QS) approach \cite{quasistatio, quasistatio_how_to} (see SM~\footnotemark[1]).
The numerical simulations on structured population have been compared to an analytically treatable mean-field version of our model dynamics.
We describe the mean-field dynamics of the group strategies directly at the group level, by a coarse-grained approach neglecting  
the microscopic dynamics of decision making that involves the group members.
Let us start defining the fraction $z_q$ of groups of size $q$ having adopted strategy $\textbf{T}$: $z_q:=\frac{N_T(q)}{N(q)}$, such that $0 \leq z_q \leq 1$.
In the limit $N\rightarrow \infty$, the 
time evolution for $z_q$ is described by the following equation \cite{fokkerplanckmastereq1, fokkerplanckmastereq2} (see SM~\footnotemark[1]):
\begin{align}
\frac{dz_q}{dt}=& (1-z_q) \sum_{q'}Q(q')  z_{q'}  p_{ST}^{qq'} \nonumber \\
&- z_q \sum_{q'}Q(q') \left( 1 - z_{q'} \right) p_{TS}^{qq'}  \label{eq:time_evolution_zq}
\end{align} 
where $p^{qq'}_{ST}$ and $p^{qq'}_{TS}$ are respectively the transition probabilities from strategy $\textbf{S}$ to $\textbf{T}$ and vice versa, given a focal group of size $q$ and a reference group of size $q'$. Such transition probabilities can be expressed as functions of the transition probabilities given the group payoffs $S/q^{\zeta}$, $L/q^{\zeta}$ and $W/q^{\zeta}$~\footnotemark[2]:
$p^{qq'}_{ST}:=(1-w_p)p^{qq'}_{SL}+w_p p^{qq'}_{SW}$ and
$p^{qq'}_{TS}:=(1-w_p)p^{qq'}_{LS}+w_p p^{qq'}_{WS}$.
Under the assumption of statistical independence $P(q|T) = Q(q)$,
we get an equation for the time evolution of $z:=\sum_q Q(q)z_q$  the fraction of groups with strategy $\textbf{T}$ in the whole population:
\begin{equation} \label{eq:repl_equation}
    \frac{dz}{dt}=z(1-z) \mathbb{E}\left[p_{ST}^{q'q}- p_{TS}^{qq'}\right]
\end{equation}
where the expectation value of function  $g(q,q')$ is defined as 
$\mathbb{E}\left[ g(q,q')\right] :=\sum_{q,q'} g(q,q') Q(q)Q(q') $. 
We recognize in Eq.~\eqref{eq:repl_equation} the celebrated Replicator Equation for a pairwise zero-sum symmetric game \cite{higherorderstabilize}, with a payoff matrix defined by
$\pi(T,S) =\mathbb{E}\left[ p_{ST}^{q'q}-p_{TS}^{qq'} \right]$, which depends in this case on $z$ through $w_p(z)$. 
Hence, besides the two absorbing states $z^*=0$ and $z^*=1$, the dynamics described by Eq.~\eqref{eq:repl_equation} has a third non trivial stationary solution, which is obtained equating to zero the expectation value in Eq.~\eqref{eq:repl_equation}.
In particular, in the limit of large population $N>>1$, we can consider $q$ continuously distributed according to $Q(q)\sim q^{-\lambda}$ and replace the sums defining the expectation value with integrals.
Under the weak selection hypothesis we can write the transition probability from a generic state $i$ to $j$ as a linear function of the payoffs difference (see SM~\footnotemark[1]): $p_{i j}=\frac{1}{2}\left[ 1 + \frac{w_F}{2} \left( \pi_{j} - \pi_{i} \right) \right]$, where $-1 \leq \frac{w_F}{2} \left( \pi_{j} - \pi_{i} \right) \leq 1$. Substituting this expression in $\mathbb{E}\left[ p_{ST}^{q'q}-p_{TS}^{qq'} \right] = 0$ and using the definitions of the group payoffs and $w_p = f(\frac{\theta}{z})$, we find the nontrivial steady state:
\begin{equation} \label{eq:zth}
z^*_{th}=\frac{\theta}{f^{-1}\left( \frac{ \zeta + \lambda - 1}{ (\zeta + \lambda + \beta - 1)(T-S)r_{\rm min}^{\beta}} \right) } 
\end{equation}
We notice that the nonlinearity introduced with the exponents $\beta$, $\zeta$ and $\lambda$ brings just a scale factor $ m:= \frac{\zeta + \lambda + \beta - 1}{\zeta + \lambda -1}$ in the solution, without changing its functional form.
Since $0 \leq z^*_{th} \leq 1$ by definition,
we can define a normalized quantity $0 \leq (T-S)' \leq 1$, where $(T-S)' := (T-S)\left[m r_{\rm min}^{\beta} f(\theta)\right]$, and rewrite Eq.~\eqref{eq:zth} simply as:
$z^*_{th}=\theta / f^{-1}\left(f(\theta)/(T-S)'\right)$.
It can be shown (see SM~\footnotemark[1]) that this steady state is an evolutionary stable state \cite{ESSdefinition}, and therefore a mixed strategy Nash Equilibrium (NE), of the pairwise zero-sum game defined by Eq.~\eqref{eq:repl_equation}.
For a probability of success inversely proportional to $z$, $w_p = \theta/z$, we can solve the Cauchy problem associated with Eq.~\eqref{eq:repl_equation} to get an analytical expression for $z(t)$ which converges to $z^*_{th}$ for all the initial conditions $0<z_0<1$ (see SM~\footnotemark[1]). 
By comparing $z^*_{th}$, the analytical expression for the null-model's nontrivial steady state, to the long term behaviour of the numerical simulations on structured populations, we can evaluate how the social hypergraph's topology influences the dynamics.
\begin{figure}[htp]
    \includegraphics[width=0.5\textwidth]{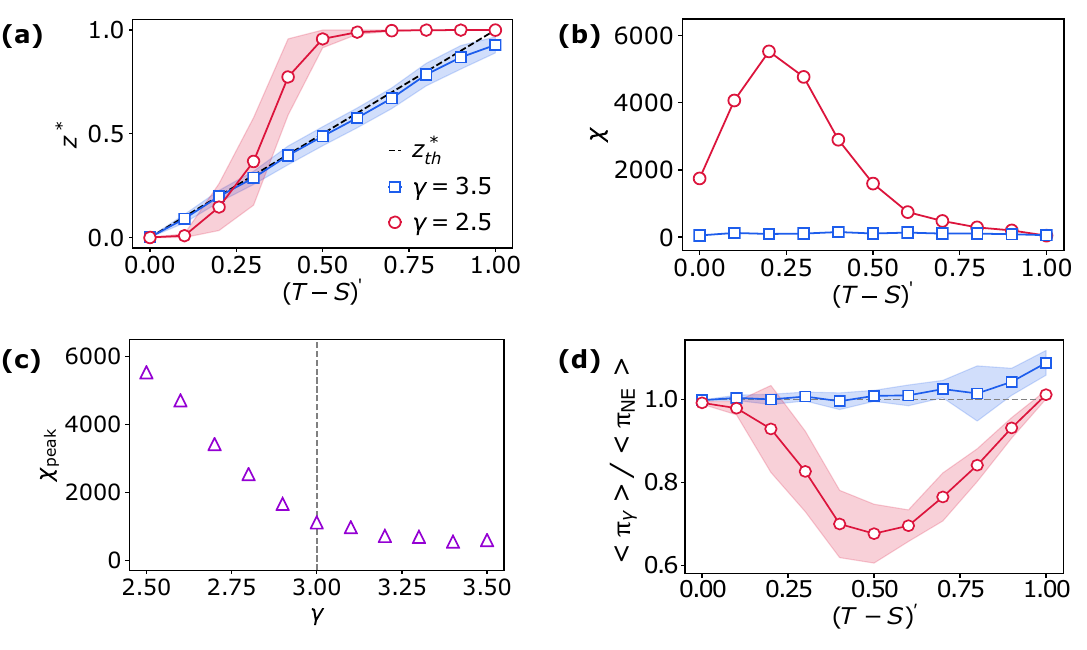}
  \caption{Long term behaviour of the simulated dynamics on hypergraphs with $H(h) \sim h^{-\gamma}$, for $w_p = \theta/z$. 
  \textbf{(a)} For $\gamma = 2.5$ we observe a phase transition, while for $\gamma = 3.5$ the QS distribution coincides with the NE (dotted line). 
  Points (circles and squares) refer to the median values of $z$.
  \textbf{(b)} Peak of the susceptibility $\chi$ at the phase transition for $\gamma = 2.5$. \textbf{(c)} Peak's height of the susceptibility as a function of the exponent $\gamma$. \textbf{(d)} Average income reduction relative to the NE payoff. 
  Shaded areas show the median absolute deviations. [Parameters: $\beta  = 0$, $\zeta = 0$, $\theta = 0.01$].
  }
\label{fig:phase_transition}
\end{figure}
Fig.~\ref{fig:phase_transition} \textbf{(a)} shows that for hypergraphs characterized by a co-membership degree distribution $H(h)$ with a finite second moment (e.g. power-law $H(h)\sim h^{-\gamma}$, with $\gamma = 3.5$)~\footnotemark[3], the numerical simulations are in good agreement with the mean-field solution in Eq.~\eqref{eq:zth}, the QS state coinciding with the NE.
Instead, when the strategy adoption dynamics takes place on heterogeneous hypergraphs with
$H(h)\sim h^{-\gamma}$, $\gamma < 3$ (results shown are for $\gamma=2.5$), we found that a phase transition occurs from the absorbing state $z = 0$ to a nontrivial stationary state $0<z<1$ that rapidly converges to $z=1$. 
To better characterize the phase transition we have computed a susceptibility function that is commonly used in SIS epidemic models \cite{pastor_sis_susceptibility}:
$
    \chi = N \frac{( \langle z^{*2} \rangle -  \langle z^* \rangle^2)}{\langle z^* \rangle}
$. 
Susceptibility functions are specifically designed to peak (diverge in the thermodynamic limit) at the critical value $(T-S)'_{c}$ of the order parameter at which the phase transition occurs. The peak of $\chi$ for $\gamma =2.5$ in Fig.~\ref{fig:phase_transition}\textbf{(b)} confirms the occurrence of a phase transition.
The plot of the maximum value of the susceptibility as a function of $\gamma$ in Fig.~\ref{fig:phase_transition}~\textbf{(c)} indicates that for $\gamma > 3$ the susceptibility is a flat function of $(T-S)'$, while for $\gamma < 3$ a peak appears, whose height increases as $\gamma$ decreases, 
pointing out $\gamma = 3$ as the threshold value for observing the phase transition.
By comparing the average group payoff in the QS state $\langle \pi_{\gamma} \rangle$ to the expected average payoff at the NE $\langle \pi_{\text{NE}} \rangle$ (see SM~\footnotemark[1]),
we see, Fig.~\ref{fig:phase_transition}~\textbf{(d)}, that the QS state for $\gamma < 3$ is sub optimal. In fact the relative average income $\langle \pi_{\gamma = 2.5} \rangle / \langle \pi_{\text{NE}} \rangle$ decreases sharply starting from $(T-S)'_{c}$ and reaches a minimum when the distance between the QS solution and the NE equilibrium is maximal (for $(T-S)'=0.5$). Instead for $\gamma > 3 $ the average payoff coincides with the NE one.  
Since the average payoff/utility decreases, the collective adoption of strategy $\textbf{T}$ observed at the phase transition can be regarded as irrational from Expected Utility Theory. 
Such irrational behaviour is due to the presence of nodes with high degree (\emph{hubs}) which, for $\gamma < 3$, implies a divergent second moment of the co-membership degree distribution \cite{newmannetwork, latora_nicosia_russo_2017}. Such hubs can trigger a strategy change in their many co-members, driving the system out of the NE.
Simulations for different values of parameters $ 0 \leq \beta, \zeta \leq 1$ show no appreciable difference with respect to the results in Fig.~\ref{fig:phase_transition}. 
However, as shown in the SM~\footnotemark[1], the synergistic parameter $\beta$ plays a role in the distribution of the average payoff as a function of the group size. 
In particular, when $\gamma < 3$, the groups of all sizes share the same loss of income for $\beta = 0$, while for $\beta > 0$ only small groups are affected by a loss of income.
Not only the average payoff, but also the strategy adoption probability $z^*_q$ shows a nontrivial dependence on the group sizes.
\begin{figure}[htp]
    \includegraphics[width=0.5\textwidth]{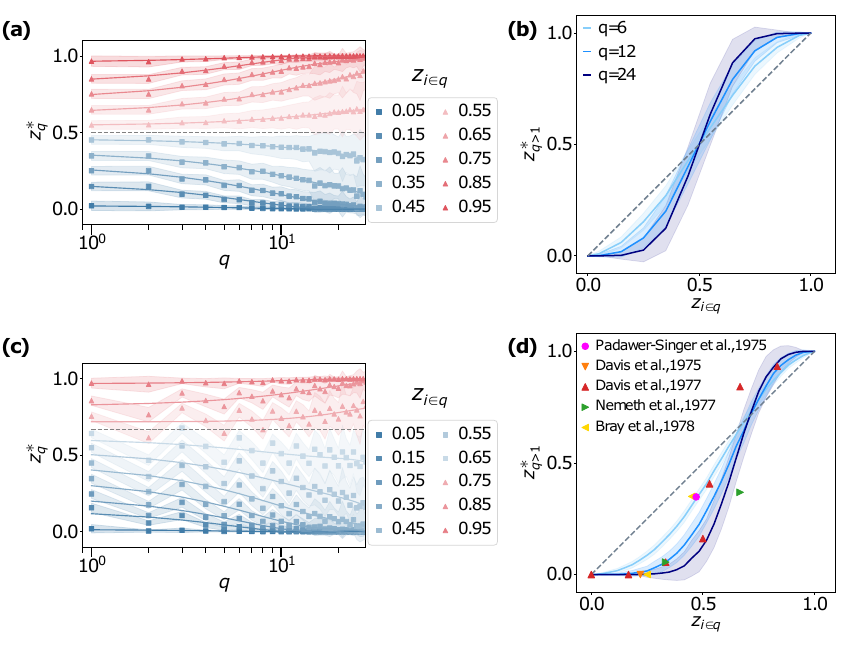}
  \caption{Choice shift under simple majority \textbf{(a)}, \textbf{(b)} and two-third majority \textbf{(c)},\textbf{(d)} decision schemes. In \textbf{(a)}, \textbf{(c)}, the simulated QS distribution $z^*_q$ as a function of group size. The grey dotted lines separate risky and safe group shifts. In \textbf{(b)}, \textbf{(d)}, $z^*_q$ for groups of size $6$, $12$ and $24$ as a function of $z_{i\in q}$, risk propensity of the members of groups of size $q$.  The colored dots in \textbf{(d)} represents empirical data on trial juries from 
  Refs. \cite{davis1975decision, DAVIS1977, penrod1980computer, Nemeth1977, bray1978authoritarianism}. Shaded areas are the standard deviations. [Parameters: $\beta = 0$, $\zeta = 0$, $\theta = 0.01$, $\lambda = 2.5$, $\nu = 2.5$].
}
\label{fig:comparison_twothird}
\end{figure}
Fig.~\ref{fig:comparison_twothird} displays $z^*_q$ obtained through numerical simulations under simple majority (panels \textbf{(a)},\textbf{(b)}) and two-third majority (panels \textbf{(c)},\textbf{(d)}) decision schemes.
In particular, in the two left panels we report $z^*_{q}$ as a function of group size $q$. The curves are level curves obtained for different values of $z_{i\in q}$, 
the fraction of decision makers in state \textbf{t} among the members of all groups of size $q$, that is the average individual risk propensity of group members.
From the definition of $z_{i\in q}$ it follows that 
, for a given level curve, if $z^*_q > z_{i\in q} \equiv z^*_{q=1}$ (or vice versa $z^*_q < z_{i\in q} \equiv z^*_{q=1}$) the average risk propensity of the groups is higher (lower) than the average individual risk propensity of their members, see SM~\footnotemark[1]. 
Thus, the data marked as triangles in panels \textbf{(a)} and \textbf{(c)} show  choice shift effects towards strategy \textbf{T} (risky shift), while squares display a shift towards the safer strategy. The transition between the risky and safe  group shift phase occurs at values of $z^*_q$ equal to the fraction of group members needed to agree the group strategy, respectively $z^*_q = 1/2$ and 
$z^*_q = 2/3$. 
However, independently from the decision scheme adopted, our model shows that the choice shift effect increases with the group size. This is somewhat counter-intuitive, since one would expect that the larger is the group the less probable are extreme collective decisions.
We compared the predictions of our model to data from empirical studies about group choice shifts observed in trial juries \cite{davis1975decision, DAVIS1977, penrod1980computer, Nemeth1977, bray1978authoritarianism} (see SM~\footnotemark[1]).
To do this, in panel \textbf{(b)} and \textbf{(d)} we plot $z^*_{q}$ as a function of $z_{i\in q}$, where the curves are level curves for different sizes $q$. The empirical data (dots), obtained with trial juries of size $6$, are better reproduced by the model with a two-third majority decision scheme. 
This is in agreement with the observation that a two-third majority 
scheme is 
spontaneously adopted by real-world trial juries \cite{DAVIS1977, Laughlin1986socialcombination}.

In conclusion, our evolutionary game model reproduces, without any ad hoc behavioral assumption, the shifts observed both at a local and a global scale in choice dilemmas, and links them to the structure of the underlying higher-order networks 
\cite{BATTISTON2020}. 
Our results can also explain how and why radical behaviours can emerge when decisions are taken in groups. Given that co-membership, group size and hyperdegree distributions are easily measurable quantities, our model  can provide useful indications on upcoming radical and potentially dangerous group behaviours in online social platforms and other real-world systems, such as financial markets.
Our work opens new paths for future research, such as the analytical characterization of the observed phase transition and the systematic exploration of the range of possible applications.

\acknowledgments
We warmly thank Lucas Lacasa and the three anonymous reviewers for their helpful comments and suggestions.


%

%
\cleardoublepage
\newpage
\onecolumngrid

\setcounter{figure}{0}
\setcounter{table}{0}
\setcounter{equation}{0}
\makeatletter
\renewcommand{\thefigure}{S\arabic{figure}}
\renewcommand{\theequation}{S\arabic{equation}}
\renewcommand{\thetable}{S\arabic{table}}
\renewcommand{\theHfigure}{S\arabic{figure}}


\section*{\large{Supplemental Material: Evolutionary game model of group choice dilemmas on hypergraphs}}
\normalsize
\vspace*{0.2 cm}

\section{Details on the hypergraphs construction and numerical simulations}

\begin{figure}[htp]
    \includegraphics[width=0.6\textwidth]{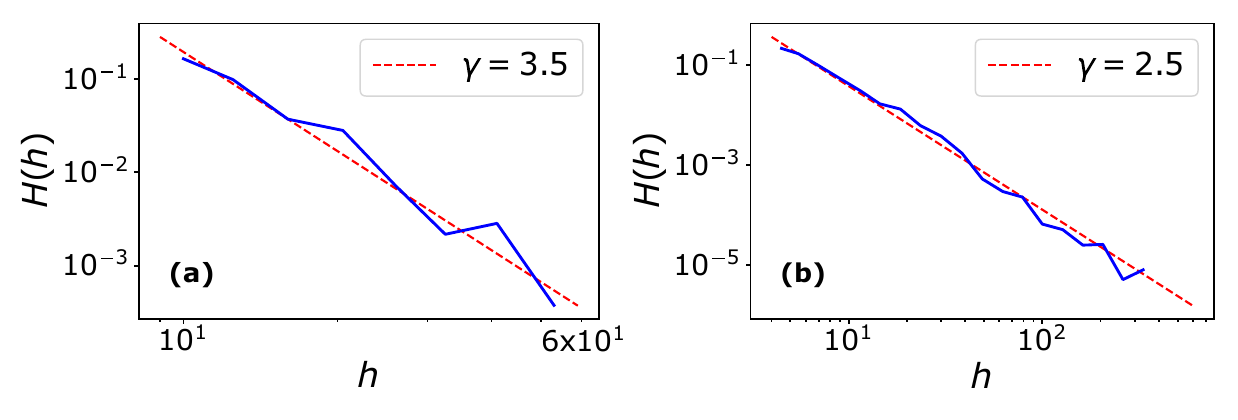}
  \caption{Long tails of the co-membership degree distribution $H(h)\sim h^{-\gamma}$ of the two hypergraphs used for the article's results. In both hypergraphs the group sizes are distributed according $Q(q)\sim q^{-\lambda}$ where $\lambda = 3.9$. The decision maker hyperdegrees instead are distributed according $K(k)\sim k^{-\nu}$ with exponents respectively equal to $\nu = 3.5$ (panel \textbf{(a)}) and $\nu = 2.5$ (panel \textbf{(b)}).
  }
\label{fig:H_distributions}
\end{figure}

The hypergraphs describing the population of agents organized in groups were built using a configuration model-like algorithm. Let us consider $M$ nodes, representing agents, and $N$ hyperedges representing groups of two or more agents. 
The algorithm works as follows. 
First, for each group member $i$ we draw its hyperdegree $k_i$ from the assigned probability distribution fuction (pdf) $K(k)$, and to each group $j$ we assign the size $q_j$, drawing it from the pdf $Q(q)$.
Then we build two separate lists, for agents and groups respectively, where the label of each agent $i$ and group $j$ is repeated  $k_i$ and $q_j$ times.
Finally, we pick uniformly at random one index $i$ from the list of agents and one index $j$ from the list of groups, and we add the index $i$ to the hyperedge $j$.
The hypergraphs built in this way will consist of different components. We restricted our analysis to the connected component with the largest number of agents.
Given the power-law distributions of the group sizes $Q(q)\sim q^{-\lambda}$ and the agent hyperdegrees $K(k)\sim k^{-\nu}$, we have checked that the resulting co-membership degree is long-tailed distributed with distribution $H(h)$. By fitting $H(h)$ with a power-law distribution (i.e. $H(h)\sim h^{-\gamma}$) we derived an exponent $\gamma$ which depends on $Q(q)$ and $K(k)$ in a non trivial way. Hence, to tune $\gamma$ we have adopted the following procedure. 
Keeping fixed $\lambda$ (i.e. the group size distribution), we changed $\nu$ in small steps checking at each step the resulting $\gamma$ by computing the co-membership degree distribution. 
In particular, the results shown in the main text correspond to two hypergraps built using $\lambda = 3.9$ and, respectively, $\nu = 3.5$ and $\nu=2.5$.
By fitting the co-membership degree distribution of the largest connected component (for the fit we used the python package \emph{powerlaw} \cite{Alstott_2014_SM}) we measured $H(h)\sim h^{-\gamma}$ with exponents respectively $\gamma \approx 3.5$ and $\gamma \approx 2.5$, as shown in Fig.~\ref{fig:H_distributions}.

All the simulations were performed on a population with a largest connected component of cardinality $M \approx 10^4$.
For our results we utilized data collected during the last $10^7$ time steps of the dynamics of $10^2$ independent runs, after a thermalisation time of $10^8$ time steps, starting from a population where  a fraction $z_0 = 0.5$ of the agents was initialized in state $\textbf{t}$, and the remaining fraction in state $\textbf{s}$.
In order to sample the quasi-stationary (QS) distribution $z^*$ and avoid the two absorbing states of the dynamics $z = 0$ and $z = 1$, we applied the method described in Refs.~\cite{quasistatio_SM, quasistatio_how_to_SM} with a memory capacity of $10^3$ states and an overwriting probability of $0.01$.
At each time step a focal agent $i$ is selected uniformly at random in the population of decision makers and one of its neighbours on the co-membership network, namely reference agent $j$, is selected with uniform probability. Then, the payoff of each of the two player is computed according the model description (see the manuscript). The focal agent can copy the state of the reference agent with a probability proportional to the payoff difference between the two individuals. If the state of the focal individual is updated, the strategy of each group in which it is involved is also updated (if there is a new majority according to decision scheme adopted by the groups).
We recall that the transition probability from the state of agent $i$ to the state of agent $j$ is usually modelled using a Fermi function: $p_{i j} = \left[1 + \exp({w_F(\pi_{i} - \pi_{j})})\right]^{-1}$, where $\pi_i$ and $\pi_j$ are respectively the payoff of the focal and reference individual. Under weak selection hypothesis (that is for small $w_F(\pi_{i} - \pi_{j})$), the Fermi function can be replaced by its linear approximation \cite{evogamefinitepop_SM,fokkerplanckmastereq2_SM}:
$p_{i j}=\frac{1}{2}\left[ 1 + \frac{w_F}{2} \left( \pi_{j} - \pi_{i} \right) \right]$, where $-1 \leq \frac{w_F}{2} \left( \pi_{j} - \pi_{i} \right) \leq 1$.
We obtained consistent results from the numerical simulations for a broad range of values of the strategies adoption strength $w_F$, both using the Fermi function and its linear approximation.
In particular, the results shown in the main text were obtained using a strategy transition probability modelled using a Fermi function with a strength of strategy selection $w_{F}$ such that $\max_{i,j}\left(\left|\pi_{j} - \pi_{i}\right|\right)w_{F}=1$, where $\max_{i,j}\left(\left|\pi_{j} - \pi_{i}\right|\right)$ is the maximum payoffs difference (in absolute value) between two agents in the population. We computed $w_{F}$ in the first simulation run of each simulations series, initializing $w_{F}$ to $1$ and then updating it every time a new maximum of the payoffs difference $\left|\pi_{j} - \pi_{i}\right|$ was encountered: $w_{F} = w_{F} / \left|\pi_{j} - \pi_{i}\right|$. This initial run was exclusively used to tune $w_{F}$, and no data were collected from it.

\section{Game for a population of two groups}

We consider a population divided into two groups of equal size $q$. As for the derivation of the stationary state $z^*$ in the main text, we describe the model at the group level using a coarse-grained approach. Since the sizes of the two groups are equal and constant (during the dynamics) both the groups cost and payoff functions are equal and constant, even if in principle they can depend on $q$. Therefore, we neglect the respective scaling factors $q^{-\beta}$ (for the cost) and $q^{-\zeta}$ (for the payoffs), defining the cost simply as $C$ and the payoffs as:
\begin{align}
\pi(\textbf{S}) &= S \\
\pi(\textbf{T}) &= \begin{cases}
            W := T-C \text{, with probability $w_p $} \nonumber \\
            L := S-C \text{, with probability $1-w_p$}
            \end{cases}
\end{align}
where $w_p = \theta N / N_T$ is the probability of earning the high reward $W$ associated to the risky strategy \textbf{T}.
In particular, for a population of $N=2$ groups, $N_T\in \{ 0, 1, 2 \}$.
For this population is then possible to write a symmetric payoffs matrix describing the pairwise game:
\begin{table}[h!]
    \setlength{\extrarowheight}{4pt}
    \begin{tabular}{cc|c|c|}
      & \multicolumn{1}{c}{} & \multicolumn{1}{c}{\textbf{T}}  & \multicolumn{1}{c}{\textbf{S}} \\\cline{3-4}
     $z$  & \textbf{T} & $\;a\;$ & $\; b\; $ \\\cline{3-4}
      $1-z$ & \textbf{S} & $S$ & $S$  \\\cline{3-4}
    \end{tabular}
    \caption{Symmetric payoffs matrix for a population of two groups of equal size. The row player represents one group, the column player the other one.}
\end{table}
\\
where $a = \theta(T-C)+(1-\theta)(S-C)$, $b = 2\theta(T-C) + (1-2\theta)(S-C)$ and $\bm{z}=(z,1-z)$ defines a mixed strategies profile (i.e. $0<z<1$ is the probability of strategy \textbf{T} adoption). Since the payoffs matrix is symmetric, we omitted the payoffs for the column player.
For $a < S$ and $S < b$, that is for $ \frac{1}{2} < \frac{\theta(T-S)}{C} < 1 $, this payoffs matrix is the payoffs matrix of the game of Chicken with ties for middle payoffs (i.e. the Volunteer's Dilemma) \cite{Bruns2015NamesFG_SM}. Since a pairwise game is completely defined by its payoffs matrix, our game for two groups and the pairwise game of Chicken are equivalent.
The game have two pure strategies Nash Equilibria (NE): (\textbf{T},\textbf{S}) and (\textbf{S},\textbf{T}). However, these pure strategies NE are asymmetric and require coordination to work (i.e. the two players deciding in advance who is going for \textbf{S} and who for \textbf{T}) \cite{archetti_2009_SM}.  The only symmetric NE of the game is $\bm{z^*} = (z^*,1-z^*)$, a mixed strategies profile NE, which can be found equalizing the payoffs on the support of $\bm{z^*}$:
\begin{equation} \label{eq:equal_support}
    z^* a + (1 - z^*)b = z^* S + (1 - z^*)S
\end{equation}
Eq.~\eqref{eq:equal_support} leads to:
\begin{equation}
    z^* = \frac{b - S}{b - a} = 2 - \frac{C}{\theta(T-S)}
\end{equation}
such that $0 < z^* < 1$ for $ \frac{1}{2} < \frac{\theta(T-S)}{C} < 1 $, which means that the mixed NE can assume any meaningful value for a probability (i.e. in range $(0,1)$) \cite{gameofchicken_1966_SM}.

\section{Derivation of Eq.~(3)}

Eq.~(3) in the main text has been derived by adapting the procedure described in Refs. \cite{fokkerplanckmastereq1_SM, fokkerplanckmastereq2_SM} to a compartmental approach.
We first write the Master Equation describing the dynamics in the sub-population (compartment) of groups of size $q$.
We recall that we are describing the mean-field dynamics for the model using a coarse-grained approach.
It is important to notice that the dynamics of the system is a Markovian process, since the strategy transition probabilities only depend on the current strategies and payoffs of the groups.
Therefore, the probability $P(N_T(q),\tau)$ of being at the time step $\tau$ in a state characterized by exactly $N_T(q)$ groups of size $q$ with strategy \textbf{T} satisfies the following equation:
\begin{align} \nonumber
    P\left(N_T(q),\tau + 1\right) - P\left(N_T(q),\tau\right) =  
    &+ P\left(N_T(q)-1,\tau\right) \sum_{q'}  T_{q'}^{+}\left(N_T(q)-1\right) \\ \nonumber
    &+ P\left(N_T(q)+1,\tau\right) \sum_{q'} T_{q'}^{-}\left(N_T(q)+1\right) \\
    &- P\left(N_T(q),\tau\right) \sum_{q'} \left( T_{q'}^{+}\left(N_T(q)\right) + T_{q'}^{-}\left(N_T(q)\right) \right)
\end{align}
Introducing the quantities $z_q Q(q) = N_T(q) / N$, $t=\tau/N$ and the probability density $\rho\left(Q(q)z_q,t\right)=N P\left(N_T(q),\tau\right)$ 
yields:
\begin{align} \nonumber
\rho\left(Q(q)z_q,t + N^{-1} \right)-\rho\left( Q(q)z_q,t \right) = 
&+\rho\left(Q(q)z_q-N^{-1},t\right)\sum_{q'}T_{q'}^{+}\left( Q(q)z_q - N^{-1} \right) \\ \nonumber
&+ \rho\left( Q(q)z_q + N^{-1} ,t \right)\sum_{q'}T_{q'}^{-}\left( Q(q)z_q + N^{-1}  \right)\\ \label{eq:master_rho}
&-\rho\left( Q(q)z_q, t \right)\sum_{q'} \left( T_{q'}^{+}\left(Q(q)z_q\right) + T_{q'}^{-}\left(Q(q)z_q\right) \right)
\end{align}
where $T_{q'}^{+}\left(N_T(q)\right)$ and $T_{q'}^{-}\left(N_T(q)\right)$ are respectively the probabilities of increasing and decreasing by one the number $N_T(q)$ of groups of size $q$ through the interaction with a group of size $q'$. 
They can be expressed as:
\begin{align}
T_{q'}^{+}\left(Q(q)z_q\right) = &\left[ Q(q) \left( 1 - z_q \right) Q(q') z_{q'}\right] p_{ST}^{qq'} \label{eq:Tplus} \\
T_{q'}^{-}\left(Q(q)z_q\right) = &\left[ Q(q) z_q Q(q') \left( 1 - z_{q'} \right) \right] p_{TS}^{qq'} \label{eq:Tminus}
\end{align}
where $z_q = N_T(q)/N(q)$ and $Q(q)$ is the probability distribution of the group sizes.
Therefore, the term in square brackets in Eq.~\eqref{eq:Tplus} represents the probability of picking at random a group of size $q$ with strategy \textbf{S} and a group of size $q'$ with strategy \textbf{T}. $p_{ST}^{qq'}$ is instead the probability of strategy transition from \textbf{S} to \textbf{T} given the group sizes, described in the main text.
Eq.~\eqref{eq:Tminus} has an analogous interpretation.
To simplify the notation, let us define $x := Q(q)z_q$ (it is worth to notice that $Q(q)$ does not change during the dynamics, therefore $x\propto z_q$).
For $N>>1$, we can perform a Kramers-Moyal expansion of the Master Equation. Using Taylor series expansions of the probability densities and the transition probabilities, and neglecting higher order terms in $N^{-1}$, we get:
\begin{align}
&\rho \left( x,t+N^{-1}\right)-\rho(x,t)\simeq \frac{d \rho(x,t)}{dt} N^{-1}
\\
&\rho \left( x \pm N^{-1},t \right) \simeq \rho(x,t) \pm  \frac{d \rho(x,t)}{dx} N^{-1} + \frac{1}{2}\frac{d^2\rho(x,t)}{dx^2}N^{-2}
\\
&T^{\pm}_{q'}(x\pm N^{-1})\simeq T^{\pm}_{q'}(x)\pm \frac{dT^{\pm}_{q'}(x)}{dx}N^{-1}+\frac{1}{2}\frac{d^2 T^{\pm}_{q'}(x)}{dx^2}N^{-2}
\end{align}
Substituting in Eq.~\eqref{eq:master_rho} and after some manipulation, we obtain:
\begin{align} \nonumber
\frac{d\rho}{dt}= &- \rho \sum_{q'} \frac{d\left( T^{+}_{q'} - T^{-}{q'}\right)}{dx} - \frac{d\rho}{dx}\sum_{q'}\left( T^{+}_{q'} - T^{-}_{q'} \right)
\\
&+\frac{d\rho}{dx}\sum_{q'}\frac{d\left( T^{+}_{q'} + T^{-}_{q'}\right)}{dx}N^{-1}
+\frac{1}{2}\left[\rho\sum_{q'}\left( \frac{d^2T^{+}_{q'}}{dx^2} +\frac{d^2T^{-}_{q'}}{dx^2}\right) + \frac{d^2\rho}{dx^2}\left( T^{+}_{q'}+T^{-}_{q'}\right)\right]N^{-1}
\end{align}
where for a matter of convenience we used a simplified notation for $\rho(x,t)$ and $T^{\pm}_{q'}(x)$, omitting the variables $x$ and $t$.
The previous equation can be written in a more convenient form as:
\begin{equation}
    \frac{d\rho}{dt} = - \frac{d}{dx}\left[ \rho \sum_{q'} \left( T^+_{q'}-T^-_{q'}\right) \right] + \frac{1}{2}\frac{d^2}{dx^2}\left[ \rho \sum_{q'}\left( T^+_{q'}+T^-_{q'} \right) \right]N^{-1}
\end{equation}
This equation is in the form of a Fokker-Plank equation:
\begin{equation}
    \frac{d\rho(x,t)}{dt}=-\frac{d\left( \rho(x,t)a(x) \right)}{dx} + \frac{1}{2} \frac{d^2\left( \rho b^2(x) \right)}{dx^2}
\end{equation}
where $a(x)= \sum_{q'} \left( T^+_{q'}(x)-T^-_{q'}(x)\right)$ is the drift and $\frac{1}{2}b^2(x) = \frac{1}{2}\sum_{q'}\left( T^+_{q'}(x)+T^-_{q'}(x)\right)N^{-1}$ is the diffusion coefficient.
Since the internal noise is microscopically uncorrelated in time, as subsequent steps of the dynamics are independent, the It\^{o} calculus applies and we obtain the Langevin equation:
\begin{equation}
    \dot{x} = a(x) + b(x)\xi
\end{equation}
where $\xi$ is uncorrelated Gaussian noise.
Taking the limit $N\rightarrow \infty$, $b(x) \propto N^{-1/2}\rightarrow 0$ and we obtain the deterministic equation:
\begin{equation}
    \dot{x}=a(x)
\end{equation}
By substituting in $a(x)$ the definition of $T^+_{q'}(x)$, $T^-_{q'}(x)$ and $x$, we finally obtain:
\begin{equation} \label{eq:zq_evo}
\frac{dz_q}{dt} = (1-z_q) \sum_{q'}Q(q')  z_{q'}  p_{ST}^{qq'} - z_q \sum_{q'}Q(q') \left( 1 - z_{q'} \right) p_{TS}^{qq'}
\end{equation} 

\section{Evolutionary stable state of the mean-field dynamics }

In order to prove that $z^*_{th}$ (such that the payoffs matrix element $\mathbb{E}\left[ p_{ST}^{q'q}-p_{TS}^{qq'} \right] = 0$) is an evolutionary stable strategy (ESS), 
we need to prove that,  $\forall z \neq z^*_{th}$, $\exists \epsilon_0(z)>0 $ such that $\forall \epsilon < \epsilon_0(z)$, we have \cite{ESSdefinition_SM}:
\begin{equation}
\pi(z^*_{th},(1-\epsilon)z^*_{th}+\epsilon z )> \pi(z,(1-\epsilon)z^*_{th} + \epsilon z) \label{eq:ESS_condition}
\end{equation}
That is, in a population where a fraction $1-\epsilon$ of the agents adopts the mixed strategy $z^*_{th}$ and a fraction $\epsilon$ the strategy $z$, if $z^*_{th}$ is an ESS there exists a threshold fraction $\epsilon_0(z^*_{th})$ below which strategy $z^*_{th}$ provides an higher expected payoff than $z$. This implies that on average every mutant strategy $z$ is eliminated by $z^*_{th}$ before reaching the critical fraction $\epsilon_0(z)$.
For convenience, let define
\begin{equation} \label{eq:zprime_def}
   z'=(1-\epsilon)z^*_{th} +\epsilon z
\end{equation}
as the fraction of groups than on average plays strategy $\textbf{T}$ (being $z$ and $z^*_{th}$ the probabilities of playing $\textbf{T}$ respectively for a fraction $\epsilon$ and $1-\epsilon$ of the groups population).
We recall that the winning probability $w_p$ that appears in the payoffs matrix element is a function of the fraction $z'$ of groups  adopting \textbf{T} in the population: $w_p=f\left(\frac{\theta }{z'}\right)$.
As a consequence the payoffs in Eq.~\eqref{eq:ESS_condition} depend implicitly on $z'$.
Let us rewrite the payoffs matrix element, by making explicit its dependence on $z'$, as:
\begin{align} \nonumber
    M_{z'} :=& \pi(T,S)|_{z'} = \mathbb{E}\left[ p_{ST}^{q'q}-p_{TS}^{qq'} \right] 
    \\ \nonumber
    =& \left[ 1 - f\left( \frac{\theta}{z'} \right) \right] R_{SL} + f\left( \frac{\theta}{z'} \right)  R_{SW}
    \\  \label{eq:M_z}
    =& R_{SL} + f\left( \frac{\theta}{z'} \right)\left[ R_{SW} - R_{SL} \right]
\end{align}
where we have used the definitions of $p_{ST}^{qq'}$ and $p_{TS}^{qq'}$ and we have introduced the quantities
\begin{align} \label{eq:RSL}
R_{SL} :=& \mathbb{E}\left[ p_{SL}^{qq'} - p_{LS}^{qq'} \right]\\ \label{eq:RSW}
R_{SW} :=& \mathbb{E}\left[ p_{SW}^{qq'} - p_{WS}^{qq'}\right] 
\end{align}
We recall that we are assuming the transition probabilities to be linear in the payoffs difference: $p_{s1,s2}^{qq'}=\frac{1}{2}\left[ 1 + w_{lin}\left( \pi_{s2} - \pi_{s1} \right)  \right]$, where $w_{lin}$ is the coefficient of linear proportionality, such that $ -1 \leq w_{lin} \left( \pi_{s2} - \pi_{s1} \right) \leq 1$.
We can therefore substitute the transition probabilities in Eqs. \eqref{eq:RSL}, \eqref{eq:RSW} with their linear expressions, to get:
\begin{align}
R_{SL} =& -\frac{w_{lin}}{2}\mathbb{E}\left[ \frac{C(q)}{q^{\zeta}} + \frac{C(q')}{{q'}^{\zeta}}\right] \\ 
R_{SW} =& \frac{w_{lin}}{2}\mathbb{E}\left[ \frac{T - S - C(q)}{q^{\zeta}} + \frac{T - S - C(q')}{{q'}^{\zeta}} \right]
\end{align}
Since $C(q)=(r_{min} q)^{-\beta} > 0$ and $T > S$, it follows that $R_{SL} < 0$ and $R_{SW} - R_{SL} > 0$.
We stress that the payoff matrix of the pairwise zero-sum symmetric game defined by the Replicator Equation is completely determined by $M_{z'}$. In fact, being the game symmetric zero-sum it implies that the two diagonal elements are $0$ an the two off-diagonal elements are respectively $M_{z'}$ and $-M_{z'}$. 
We can rewrite $z^*_{th}$ as: 
\begin{equation}
  z^*_{th} = \frac{\theta}{f^{-1}\left(\frac{-R_{SL} }{  R_{SW} -  R_{SL} } \right)}  
\end{equation}
We can now substitute in Eq.~\eqref{eq:ESS_condition} the explicit expressions for the expected payoffs given the mixed strategy profiles $(z^*_{th},z')$ and $(z,z')$, where $z'=(1-\epsilon)z^*_{th}+\epsilon z$, obtaining:
\begin{equation}
    z^*_{th}(1-z')M_{z'} - (1-z^*_{th})z'M_{z'} > z(1-z')M_{z'}-(1-z)z'M_{z'}
\end{equation}
Bringing all the terms on the left of the inequality, and after some simple manipulations, one obtains the condition:
\begin{equation} \label{eq:ESS_brief}
    (z^*_{th} - z)M_{z'}>0
\end{equation}
It is easy to prove that this condition holds true $\forall z\neq z^*_{th}$ and $\forall \epsilon < \epsilon_0(z) \equiv 1 $ and hence that $z^*_{th} $ is an ESS.
In fact, given the definition of $M_{z'}$ in Eq.~\eqref{eq:M_z}, it follows immediately that $M_{z'}>0$ if:
\begin{equation}
    z'<\frac{\theta}{f^{-1}\left( \frac{-R_{SL}}{R_{SW}-R_{SL}} \right)} \equiv z^*_{th}
\end{equation}
Substituting the definition of $z'$,  Eq.~\eqref{eq:zprime_def}, we obtain:
\begin{equation} \label{eq:Mzpositive}
    (1-\epsilon)z < (1-\epsilon)z^*_{th}
\end{equation}
which for $\epsilon < 1$ implies $M_{z'}>0$ for
$z < z^*_{th}$ (i.e. if $z^*_{th} - z > 0$), and $M_{z'}<0$ for $z^*_{th} - z < 0$.
Hence, if $z^*_{th}-z>0$, then $\forall \epsilon < 1 \equiv \epsilon_0$ we have $M_{z'}>0$, and the condition Eq.~\eqref{eq:ESS_brief} holds true. 
On the other hand, if $z^*_{th}-z<0$ from Eq.~\eqref{eq:Mzpositive} it follows that also $M_{z'}<0$ $\forall \epsilon < 1 \equiv \epsilon_0$, and Eq.~\eqref{eq:ESS_brief} is again satisfied. 
\\
Thus, $z^*_{th}$ is an ESS.

\section{An analytical expression for \texorpdfstring{$\mathbf{z(t)}$}{TEXT}}

\begin{figure}[htp]
    \includegraphics[width=0.4\textwidth]{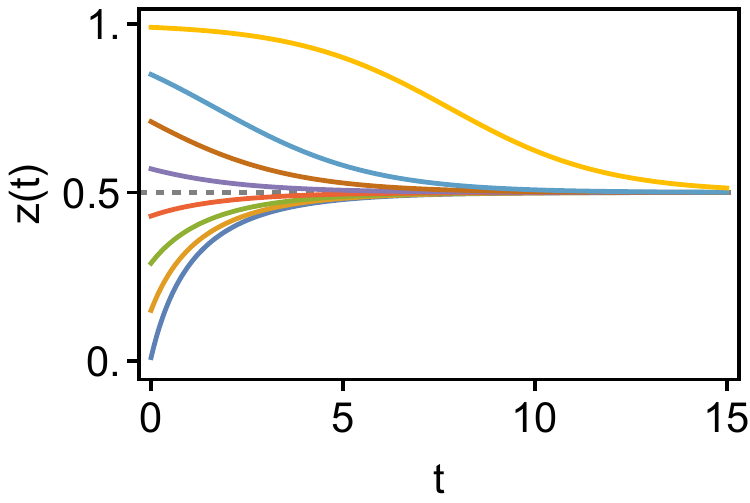}
  \caption{Trajectory $z(t)$ for different initial conditions $z_0$. The dotted grey line represents the attractor of the dynamics $z^* = a/b$, for $a=0.5$ and $b=1$.}
\label{fig:trajectory}
\end{figure}

The time evolution of $z$, the fraction of groups with strategy $T$, for $N>>1$ is described by the deterministic equation:
\begin{equation} \label{eq:repl_equation_sup}
    \frac{dz}{dt}=z(1-z) \mathbb{E}\left[p_{ST}^{q'q}- p_{TS}^{qq'}\right]
\end{equation}
In the particular case $w_p = \frac{\theta}{z}$ (i.e. $f(x) = x$), Eq.~\eqref{eq:repl_equation_sup} takes the form:
\begin{equation} \label{eq:repl_eq_linear}
    \frac{dz}{dt} = z(1-z)\left[\frac{a}{z} - b\right] ,
\end{equation} 
where $b>a>0$, being in our model $a=\theta\left(R_{SW} - R_{SL}\right)$ and $b=-R_{SL}$ (see the definitions of $R_{SL}$ and $R_{SW}$ in Eq.s~\eqref{eq:RSL},\eqref{eq:RSW}) .
Solving Eq.~\eqref{eq:repl_eq_linear} with initial condition $0<z_{0}:=z(0)<1$, we find:
\begin{equation}
    z(t)=\frac{a(1-z_0)-(a-b z_0)e^{t(a-b)}}{b(1 - z_0) -(a - b z_0)e^{t(a-b)}} 
\end{equation}
Since $a<b$, in the limit $t\rightarrow \infty$ the trajectory converges to $z^* = a/b$, which for our model coincides with the stationary state found in the main text: \begin{equation} \label{eq:zth_sup}
z^*_{th}=\frac{\theta}{f^{-1}\left( \frac{ \zeta + \lambda - 1}{ (\zeta + \lambda + \beta - 1)(T-S)r_{\rm min}^{\beta}} \right) } 
\end{equation}
This happens for all initial conditions $0 < z_0 < 1$, therefore $z^*_{th}$ is a global attractor of the dynamics.
Fig.~\ref{fig:trajectory} shows the trajectory $z(t)$ for different values of the initial condition $z_0$. 

\section{Nash Equilibrium (NE) expected payoff}

The expected payoff for a group of size $q$ is:
\begin{equation} \label{eq:expected_NEpayoff_forsize}
    \langle \pi_{NE} \rangle_q = z_q\left[ w_p W'_q + (1-w_p)L'_q \right] + (1-z_q)S'_q
\end{equation}
where $w_p = f(\frac{\theta}{z})$, 
$S'_q = \frac{S}{q^{\zeta}}$, 
$W'_q = \frac{W}{q^{\zeta}}$ and 
$L'_q = \frac{L}{q^{\zeta}}$.

Under the assumption of statistical independence $P(T|q)=P(T)$, we can approximate $z_q\sim z$ in Eq.~\eqref{eq:expected_NEpayoff_forsize}. Replacing $z$ with the NE solution Eq.~\eqref{eq:zth_sup}, we find the analytical expression for the expected payoff at the NE as a function of group size $q$.
\begin{equation} \label{eq:NE_payoff_size}
    \langle \pi_{NE} \rangle_{q} = \left[ \frac{\theta r_{min}^{\beta}}{f^{-1}\left(\frac{f(\theta)}{(T-S)'}\right)} \left( \frac{1}{m} - \frac{1}{q^{\beta}} \right) + S \right] \frac{1}{q^{\zeta}} 
\end{equation}
We can also compute the expected average payoff in the whole population by multiplying this expression by the group sizes distribution $Q(q)\sim q^{-\lambda}$ and integrating over $q$:
\begin{equation}
    \langle \pi_{NE} \rangle  =   \left[  \frac{\theta r_{min}^{\beta}}{f^{-1}\left(\frac{f(\theta)}{(T-S)'}\right)} \left( \frac{1}{m\left( \lambda + \zeta - 1 \right)} - \frac{1}{\lambda + \zeta + \beta - 1} \right)  +  \frac{S}{\lambda + \zeta - 1} \right] (\lambda - 1)
\end{equation}

\section{Average group payoff as a function of the group size}

\begin{figure}[htp]
    \includegraphics[width=0.8\textwidth]{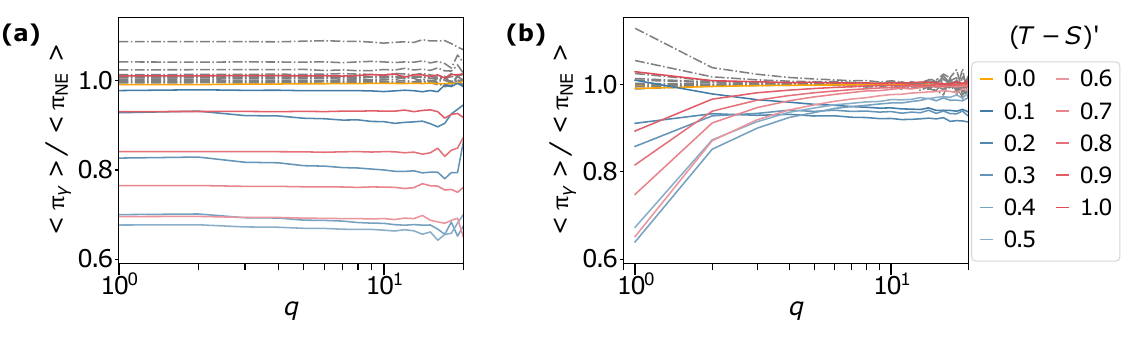}
  \caption{Average group income divided by the expected NE payoff as a function of the group size (Eq. \eqref{eq:NE_payoff_size}), for values of $\beta = 0$ (panel \textbf{(a)}) and $\beta > 0$ (panel \textbf{(b)}) and for $\gamma > 3$ (dotted lines) and $\gamma < 3$ (continuous lines). Each line refers to a different value of $ 0 \leq (T-S)' \leq 1$ used in the numerical simulations. We set the parameter $\zeta = 0$ for both the panel.}
\label{fig:ave_payoff_size}
\end{figure}

The synergistic parameter $\beta$ plays a role in the distribution of the average group payoff as a function of the group size. Fig.~\ref{fig:ave_payoff_size} shows the measured average payoff as a function of the group size $q$ divided by the expected payoff at the NE as a function of $q$ described by Eq. \eqref{eq:NE_payoff_size}. Panel \textbf{(a)} shows the results for synergistic parameter $\beta = 0$ and panel \textbf{(b)} for $\beta = 1$. It is worth to point out that for the results in Fig.~\ref{fig:ave_payoff_size} we set the parameter $\zeta=0$, in order to focus only on the role played by $\beta$.
In both panels, the continuous lines refer to hypergraphs with a scale-free co-membership structure, i.e. with $\gamma < 3$ (in particular $\gamma = 2.5$ for these results), while the dotted grey lines refer to hypergraphs with $\gamma > 3$ (in particular $\gamma = 3.5$).
We measured the group payoffs on the same hypergraphs used for the results in Fig.1 of the main manuscript, built according the algorithm described in the first section of the SM, and following the same procedure: after a thermalization time of $10^8$ simulations step, we averaged the groups payoffs as a function of the group size $q$ over the last $10^7$ simulation steps of $100$ independent simulation runs.
As Fig.~\ref{fig:ave_payoff_size} displays, when $\gamma < 3$ (continuous lines), the groups of all sizes share the same loss of income for $\beta = 0$, while for $\beta > 0$ only small groups are affected by a loss of income. 
The non-monotone  nature of the curves (they decrease till $(T-S)' \sim 0.5$ and then start going back to the NE value) is related to the behaviour of $z^*_{\gamma}$. In particular for $\gamma<3$ (as shown in Fig.1 panel \textbf{(d)} in the main manuscript), when $z^*$ reaches a maximum distance from the NE (for $(T-S)' \sim 0.5$) also the payoff reduction respect to the NE reach a maximum and then decreases as the distance between $z^*$ and the NE decreases.
If instead we focus on a specific curve in Fig.\ref{fig:ave_payoff_size} (i.e. on a curve for a specific value of $(T-S)'$), for $\beta > 0$ (Fig. \ref{fig:ave_payoff_size}\textbf{(b)}) we expect the average payoff at the NE to increase in the group size, since the cost decreases as $C(q) \sim q^{-\beta}$, as shown by Eq.\eqref{eq:NE_payoff_size}. However the payoff reduction measured in the structured populations with $\gamma<3$ does not agree with this prediction: Fig.\ref{fig:ave_payoff_size}\textbf{(b)} shows that for a synergistic factor $\beta > 0$ the measured average payoff is usually much lower than the predicted NE payoff for the small groups (with a size $q\leq 4$), while the measured and expected payoffs are in good agreement for larger groups (being the ratio between the two quantities close to $1$ for larger $q$).
A final note: the reason why we obtain essentially identical average payoff loss ( as the one shown in Fig.1 \textbf{(d)} in the main manuscript) for both $\beta = 0$ and $\beta>0$, it is that in networks with power-law distributed group sizes, the great majority of groups has small size ($q\leq 4$), so for $\beta>0$ the few groups that still earns as at the NE (i.e. the larger groups) do not influence in practice the payoff reduction observed averaging on the whole population.

\section{Mock trial juries data set}

The group strategy in our model is determined according to a social decision scheme which aggregates the individual preferences of the group members. The impact of the decision scheme over the group consensus has been investigated in a series of classic empirical studies over small and isolated groups of decision makers involved in a great variety of binary choices: from simple risk-taking tasks (typically choosing on which one of two lotteries to place a bet) to mock juries (choosing between guilty and not-guilty verdict). In particular, in the context of trial juries, as observed in Ref.\cite{kalven1966_americanjury_SM}: \emph{“to a substantial degree the jury verdict is determined by the posture of the vote at the start of the deliberation process and not by the impact of this process as rational persuasion. The jury tends to decide in the end whichever way the initial majority lies. . . . On this view the study can be thought of as a study of the sentiments that will lead to initial majorities . . . the deliberation process although rich in human interest and color appears not to be at the heart of jury decision-making.
Rather, deliberation is the route by which small group pressures produce consensus out of the initial
majority"}. 
This in practice means that, despite the different nature of the decision problem, the decisional process structure remains the same for a large variety of group decisions, included risk-taking scenarios and mock juries deliberations: a rational decision maker faces an alternative between two qualitatively different choices under uncertainty (the decision maker does not know which choice is the “right” one) and the strategy is determined in groups according to some majoritarian decision scheme. The specific decision scheme adopted has been empirically observed to depend on the specific nature of the decision task.
In particular, as pointed out by Ref.\cite{Laughlin1986socialcombination_SM} mock juries seems to adopt a two-third majority social decision scheme, where \emph{juries without a majority either are unable to come to a verdict (“hang”) or give the defendant the benefit of the doubt and acquit}.
From a practical point of view, the reason we used data from mock trial juries is simply their relative abundance (and consistency in the experimental methodology adopted to collect them) in literature. 
Fun fact: from an historical point of view, this unusual abundance of studies on trial juries conducted in the Seventies was triggered by the debate on the optimal size of a trail jury ($6$ or $12$ members) and on the best decision schemes to adopt in the US courtroom justice system started with a series of controversial jury verdicts \cite{davis1975decision_SM}.
In these empirical studies, a group of volunteers was divided in mock trial juries (most commonly of size $6$) and asked to deliberate over a real juridical case and to agree over a verdict (guilty/not guilty). The data we used are the fraction of volunteers with a personal preference over the guilty verdict before the jury deliberation and the fraction of juries (i.e. groups of decision makers) which end up with a guilty verdict.
It is worth to notice that, due the way in which data were aggregated, most of the empirical studies provided just few data points (i.e. $\mathbf{P_{G}^{ind}}$, the fraction of individuals with a pre-deliberation guilty preference in the whole population of volunteers, and $\mathbf{P_{G}^{group}}$, the corresponding fraction of mock juries with a guilty verdict).
Table \ref{table:data} shows the data from Refs.\cite{davis1975decision_SM, DAVIS1977_SM, Nemeth1977_SM, bray1978authoritarianism_SM, penrod1980computer_SM}  used in Fig.2 of the manuscript.
\begin{table}[h!]
    \setlength{\extrarowheight}{4pt}
    \begin{tabular}{c|c|c|c|c|c|c|c|c|c|c|c|c|c|}
    \cline{2-14}
      & \multicolumn{1}{c|}{\textbf{PadawerSinger1975}} & \multicolumn{1}{c|}{\textbf{Davis1975}}  & \multicolumn{7}{c|}{\textbf{Davis1977}} & \multicolumn{2}{c|}{\textbf{Nemeth1977}} & \multicolumn{2}{c|}{\textbf{Bray1978}} \\\cline{2-14}
     $\mathbf{P_{G}^{ind}}$  & $0.47$ & $0.22$ & $0.84$ & $0.67$ & $0.50$ & $0.34$ & $0.17$ & $0.00$ & $0.53$ & $0.34$ & $0.67$ & $0.25$ & $0.45$  \\
     $\mathbf{P_{G}^{group}}$  & $0.35$ & $0.00$ & $0.94$ & $0.84$ & $0.16$ & $0.06$ & $0.00$ & $0.00$ & $0.41$ & $0.06$ & $0.37$ & $0.00$ & $0.35$  \\\cline{2-14}
    \end{tabular}
    \caption{$\mathbf{P_{G}^{ind}}$, the fraction of individuals with a pre-deliberation guilty preference in the whole population of volunteers, and $\mathbf{P_{G}^{group}}$, the corresponding fraction of mock juries of size $6$ with a guilty verdict.}
    \label{table:data}
\end{table}

\section{Level curves}

The level curves in Fig.2 of the main manuscript show
$z^*_{q}$, the fraction of groups with strategy $T$ and size $q$, corresponding to $z_{i\in q}$, a given fraction of group members in state \textbf{t} among all groups of size $q$.
From a computational point of view, after waiting a thermalization time of $10^7$ simulation steps, we measured $z^*_{q}$ as a function of $z_{i\in q}$ and $q$ over the last $10^5$ simulation steps of $100$ independent simulation runs. Every $10$ simulation runs we progressively increased $(T-S)'$ (starting from $(T-S)' = 0$) in order to obtain different fraction of groups with strategy $T$ in the quasi-stationary state.
The data points in Fig.2 panels \textbf{(a)}  and \textbf{(c)} of the manuscript have been then obtained by binning $z^*_{q}$ in a given number of equally spaced bins of the corresponding $z_{i\in q}$, and averaging $z^*_{q}$ over each bin. The shaded areas represent the variance of $z^*_{q}$ in each bin.
Finally, the curves in Fig.2 panel \textbf{(a)} and \textbf{(c)} have been obtained through a 1-D smoothing spline fit of the data points from the numerical simulations.
Since the groups of size $q = 1$ are composed by a single group member, the fraction of groups of size $q=1$ with strategy $T$ coincides with the fraction of their members in state \textbf{t}, i.e. $z^*_{q=1} \equiv z_{i\in q=1}$. All the data points in a given level curve are binned according the same values of $z_{i\in q}$, that is  $z_{i\in q} \equiv z_{i\in q=1} \equiv z^*_{q=1}$, $\forall q > 1$ for a given level curve, therefore we can compare the group risk propensity $z^*_{q}$ to the individual risk propensity of group members $z_{i\in q}$ simply comparing it to $ z^*_{q=1}$ instead. This means that for a given level curves in Fig.2 panels \textbf{(a)} and \textbf{(c)}, if $z^*_{q} > z^*_{q=1} \equiv z_{i\in q}$ the group risk propensity of the groups of size $q$ is greater than the risk propensity of their group members and we have a risky shift, vice versa if $z^*_{q} < z^*_{q=1} \equiv z_{i\in q}$ we have a safe shift.
It is worth pointing out that the oscillating convergence that can be observed in Fig.2\textbf{(c)} is a consequence of the decision scheme and of the discrete nature and the finite size of groups. For example, if a $2/3$ majority of the group members are required to agree a decision, this obviously translates into different “theoretical” majorities depending on the group size: e.g. for groups of size $6$ $\rightarrow$ $6*2/3 = 4$ (i.e. $4$ group members are needed to agree a group strategy), for size $7$ $\rightarrow$ $7*2/3 = 4.6$, for size $8$ $\rightarrow$ $8*2/3 = 5.3$, etc... But the groups are discrete object, therefore in practice we have to “roof” these quantities: in a group of size $7$ at least $5$ group members are effectively required instead of the theoretical $4.6$ (there is no such a thing as $0.6$ group member), with an approximation of $0.4$; for a group of size $8$ the “effective” threshold for a majority is $6$ group members, and in this case we are approximating of $0.7$ respect the theoretical $5.3$ group members required for a $2/3$ majority, etc.. Therefore for different sizes, we have different amount of approximation due to the discrete and finite nature of groups. Hence, the oscillation observed in Fig.3 (now corresponding to Fig.2 in the revised version of the manuscript) is due to the periodicity of the remainder of the operation of division between two integer numbers (i.e. the size of the groups multiplied by the given fraction of group members needed to agree a decision). This behaviour can be observed, although it is less obvious, also for the simple majority decision scheme Fig.2\textbf{(a)} (where instead of a $2/3$ majority we have a $1/2$ majority), with a periodicity (over the group sizes) equal $2$.

\end{document}